\documentclass[traditabstract]{aa}  

\usepackage{natbib}
\bibpunct{(}{)}{;}{a}{}{,} 
\usepackage{graphicx}
\usepackage{txfonts}
\def\note #1]{\noindent{\bf #1]}}

\def\sgnk2{{\rm sgn\left(K^2\right)}}
\def\rmd{{\rm d}}

\begin{document}
   \title{Probing tiny convective cores with the acoustic modes of lowest degree}

   \subtitle{}

   \author{M.~S.~Cunha
          \and
          I.~M.~ Brand\~ao
          }

   \offprints{M.S. Cunha}

   \institute{Centro de Astrof{\'\i}sica e Faculdade de Ci\^encias, Universidade do Porto, Rua das Estrelas, 4150-Porto, Portugal\\
              \email{mcunha@astro.up.pt}
             }

   \date{}

   \abstract{ Solar-like oscillations are expected to be excited in stars of up to about 1.6 solar masses. Most of these stars will have convective cores during their Main-sequence evolution. At the edges of these convective cores there is a rapid variation in the sound speed which influences the frequencies of acoustic oscillations.
In this paper we build on earlier work by Cunha and Metcalfe, to investigate further the impact that these rapid structural variations have on different p-mode frequency combinations, involving modes of low degree. In particular, we adopt a different expression to describe the sound speed variation at the edge of the core, which we show to reproduce more closely the profiles derived from the equilibrium models. We analyse the impact of this change on the frequency perturbation derived for radial modes. 
Moreover, we consider three different small frequency separations involving, respectively, modes of degree $l=0,1,2,3$, $l=0,1$, and $l=0,2$, and show that they are all significantly affected by the sharp sound speed variation at the edge of the core. In particular, we confirm that the frequency derivative of the diagnostic tool that combines modes of degree up to 3 can potentially be used to infer directly the amplitude of the relative sound speed variation at the edge of the core. Concerning the other two diagnostic tools, we show that at high frequencies they can be up to a few $\mu$Hz smaller than what would be expected in the absence of the rapid structural variation at the edge of the core. Also, we show that the absolute values of their frequency derivatives are significantly increased, in a manner that is strongly dependent on stellar age. 
   
 }

   \maketitle
%

\section{Introduction}

Asteroseismology is a fast developing research subject, with potential to contribute greatly to our understanding of the physics of stellar interiors and stellar evolution \cite[e.g.][for a review]{cunhaetal07}. One of the benchmarks of asteroseismology was undoubtedly the detection of oscillations in solar-like pulsators \cite[e.g.][and references thereafter]{bedding07}, which was made possible by the development of a new generation of spectrographs with high enough precision to detect the small amplitudes (typically of a few tens of cm/s) of stochastically-driven oscillations, in stars other than the sun. Since then, observations of pulsating stars from space have also become possible, first with the tiny 5~cm star tracker mounted on the NASA WIRE satellite \citep{buzasietal00}, followed by the 15~cm Canadian-led satellite MOST \citep{walkeretal03}, and the 27~cm French-led satellite CoRoT \citep{micheletal08}. These satellites and high-precision ground-based campaigns conducted in the past few years \citep[e.g.][]{arentoftetal08}, have allowed the study of a number of bright solar-type stars. Yet, the recent launch of the NASA Kepler satellite \citep{gillilandetal10} is expected to completely change the current picture, increasing to over a thousand the number of solar-like pulsators for which excellent seismic data sets will be available. Moreover, for a subset of these, Kepler will provide very long (up to 3.5 yr) time-series of data, which for sure will offer an unprecedented insight on solar-like pulsators.

The optimal exploitation of the seismic data on solar-like pulsators requires that a great effort is put also into developing seismic diagnostic tools that may be used to infer information about the internal structure of these stars. In particular, making inferences about their deepest layers and, consequently, improving the description of physical phenomena such as diffusion and convective overshoot in stellar evolution codes, is the only way to improve the age and mass determinations that are derived in asteroseismic studies, and that are so valuable in a more general context of astrophysics.

With this aim in mind, \citet[][hereafter paper~I]{cunha07} have investigated on the possibility of making inferences about the properties of the tiny convective cores present in most main-sequence solar-like pulsators. The immediate goal of their study was two-fold: firstly, to derive the perturbation that the abrupt change in sound speed at the edge of the convective core produces on the oscillation frequencies of radial modes; secondly, to build a diagnostic tool - i.e. a combination of oscillation frequencies - capable of isolating that perturbation. 

In the past, other authors have predicted the signal produced in the oscillation frequencies by sharp structural changes in the interior of stars, either associated with edges of outer convective regions, or with regions of element ionization \citep{monteiroetal94,monteiroetal00,monteiro05,houdek07}. However, with the exception of the work of \cite{houdek07}, their analyses assumed that the sharp structural variations took place well inside the propagation cavity of the waves, a condition that is not fulfilled when considering convective cores as small as those present in main-sequence solar-like pulsators. Studies of the signal produced by sharp variations taking place specifically at the edge of stellar or planet cores, different from that presented in Paper~I, have also been carried out in the past \citep{provost93,roxburgh99,roxburgh01,roxburgh07,miglio08}. Of the latter, only the work of \cite{roxburgh07} was directed towards the study of main-sequence solar-like pulsators.

In paper~I the authors considered an asymptotic formulation for the eigenfunctions that is valid in the region of the turning point of the wave equation. The main conclusions of their work were: (1) that the sharp variation of the sound speed at the edge of the tiny convective core produces a decrease in the oscillation frequencies of high radial order modes, the absolute value of which increases with frequency; (2) that a particular combination of modes of degrees $l=0,1,2,3$ can be used to recover, in the asymptotic regime of high radial orders, the frequency perturbations, except for a proportionality constant that is independent of stellar age; (3) that this diagnostic tool is sensitive to the size of the convective core, the star's age, and the stage at which the core begins to contract.

In the present work we follow on the analysis presented in paper~I, but consider two additional aspects which are of particular relevance when anticipating the application to real data. Firstly, in the derivation of the frequency perturbations due to the sharp structural change at the edge of the core, we improve the modulation of the sound speed perturbation, which in the earlier work was considered to be a simple step function. This improvement and its implication for the results will be discussed in sections \ref{sec:deltac} and \ref{sec:conclusions}. Secondly, given the well known difficulty in detecting modes of $l=3$ in space-based data, we analyse the impact that the derived frequency perturbations have on seismic diagnostic tools that involve only modes of degree $l\le 2$. This analysis will be presented in section \ref{sec:quantities} and further discussed in section \ref{sec:conclusions}.

\section{Signature of a small convective core\label{sec:deltac}}

\subsection{The sound speed perturbation}

There is still no consensus about the exact mass above which a main-sequence star may have a convective core. Nevertheless, solar-like pulsators more massive than about 1.1~M${_\odot}$ (where M${_\odot}$ is the mass of the sun) are expected to develop convective cores at some point during their main-sequence evolution. Moreover, the size of the core increases as the star evolves from the Zero Age Main Sequence (ZAMS) and later decreases again, as it approaches the Terminal Age Main Sequence (TAMS). For a 1.3~M${_\odot}$ star, for instance, the convective core reaches a maximum size of about $9\%$ of the stellar radius, at about half way through its life in the main-sequence, the exact size depending on the amount of overshooting considered. 

The abrupt change in the chemical composition, hence, in stellar density, at the edge of the convective core, results in a corresponding abrupt change in the sound speed profile. The shape of the sound speed profile at the edge of the core is expected to be influenced by local physical processes, such as diffusion and semi-convection. Thus, by characterizing the former it is in principle possible to infer about these physical processes, as well as about more general properties of the core, such as its size.

\begin{figure}
   \centering
   \includegraphics[width=8cm]{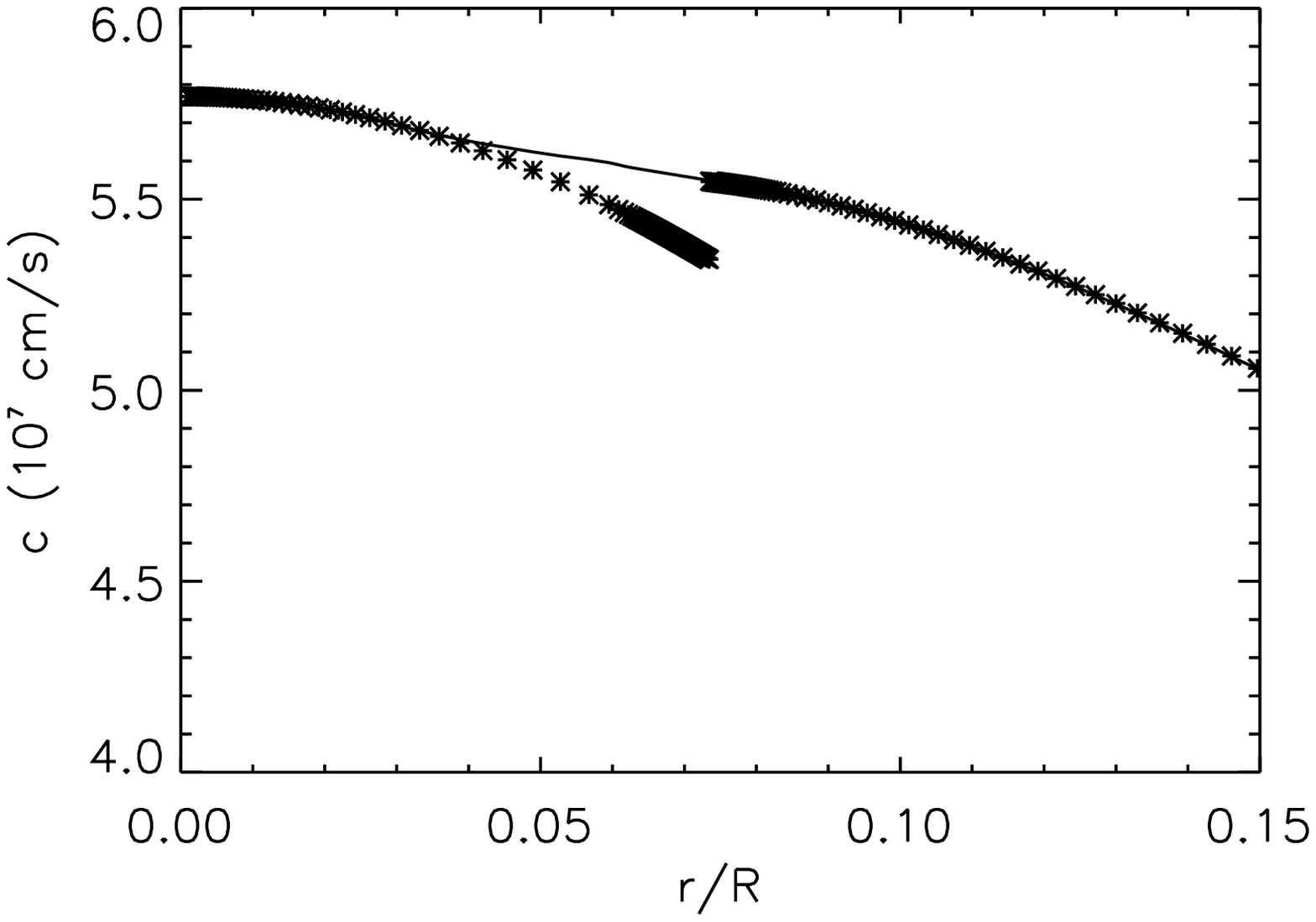}
   \includegraphics[width=8cm]{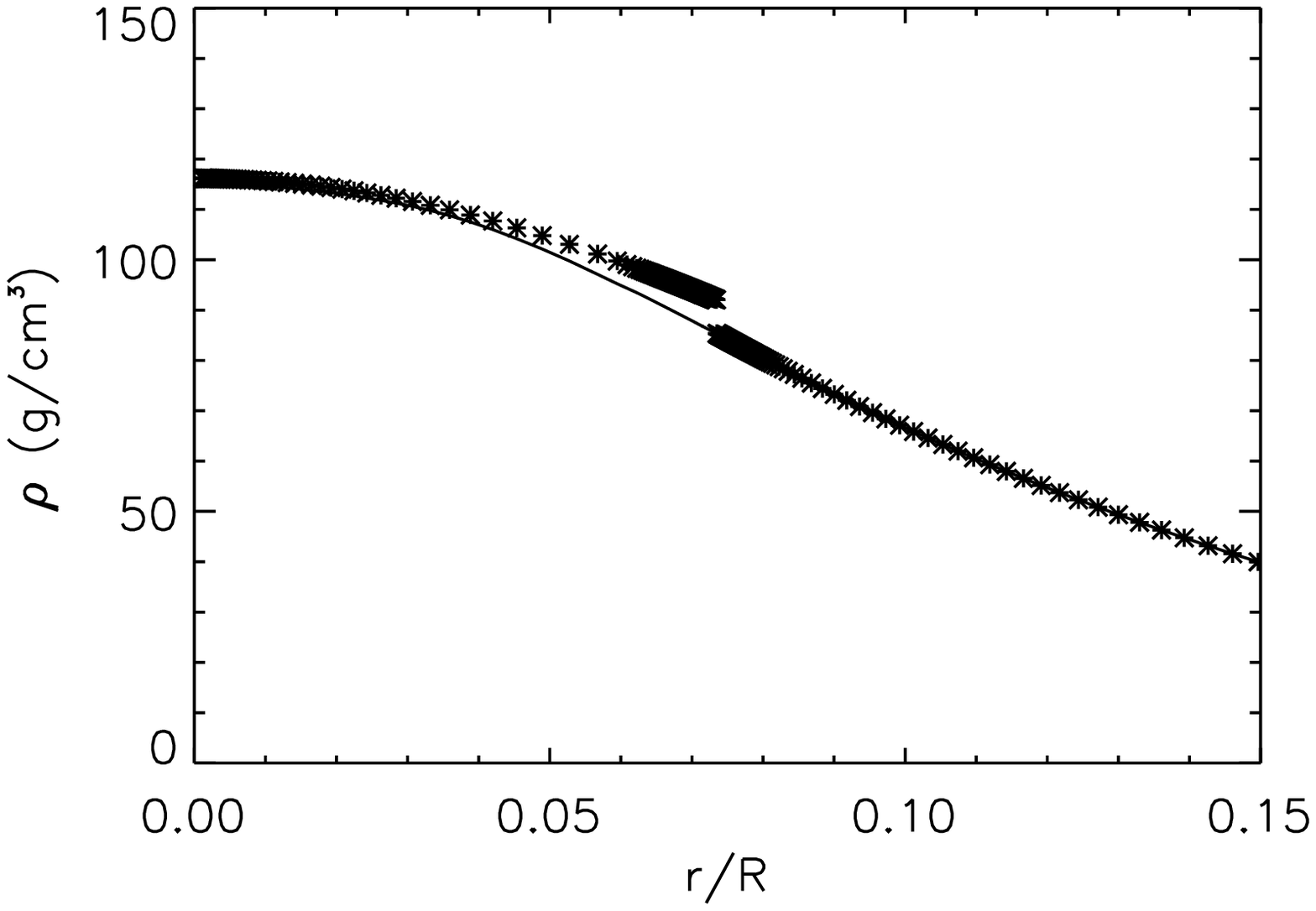}
      \caption{Sound speed (upper panel) and density (lower panel) profiles for  a model with $M=$1.3~M${_\odot}$ and an age of 2.25 Gyr. The symbols show the profiles in the original model, computed with the evolution code ASTEC, while the continuous lines show the profiles in the smooth model, obtained through the iterative procedure described in the text. }
         \label{cdens}
   \end{figure}

To derive an expression for the signature of the convective core on the oscillation frequencies, we need to model the sound speed sharp variation in terms of a perturbation to the sound speed profile that would be found in an otherwise similar model, with a smooth structure at the edge of the core. The smooth model is obtained from the original model, by an iterative procedure. First, the sound speed profile around the edge of the core is smoothed arbitrarily, by means of low order polynomial fit between the two (also arbitrary) ends of the perturbation region. Then the density profile is recomputed assuming hydrostatic equilibrium and compared with the density profile of the original model. Generally speaking, the new density profile will differ from the original one all the way from the center of the star to the radius at which the smoothing of the sound speed started. Since we want the two models to differ only around the edge of the core (i.e., the perturbation region), we then iterate, by modifying the form of the smoothed sound speed profile, until both the sound speed profile and the density profile in the original and smoothed models differ only in a limited region around the edge of the core. Figure \ref{cdens} shows a comparison of the smoothed and original sound speeds and density profiles, in a model of a 1.3~M${_\odot}$ with an age of 2.25~Gyr, computed with the evolution code ASTEC \citep{cd08} with a core overshoot of $\alpha_{ov}=0.25$ and no diffusion. All models used in this work are part of a 1.3~M${_\odot}$ evolution sequence -- the same as the one used in paper~I. Moreover, the pulsation properties of these models, such as the eigenfrequencies and eigenfunctions, that shall be used later, were computed with the pulsation code ADIPLS \citep{cd08}

In order to compute the frequency perturbation associated with the sharp variation of the sound speed, we need to have an expression for the relative perturbation to the sound speed squared, $\delta c^2/c^2$. In paper~I that perturbation was described by a step function. Here we consider a different analytical expression for the perturbation, more in line with what is expected from the inspection of the stellar models. Figure~\ref{c2} shows $\delta c^2/c^2$ for two stellar models, both with mass $M=$ 1.3~M${_\odot}$ and with ages of 2.25~Gyr and 4.0~Gyr, respectively. These were computed directly from the original and smooth models, the difference being in the sense original-smooth model. Clearly, a simple step function will not reproduce the smooth decline of the perturbation at smaller values of the radius. Consequently, here we adopt, instead, a two parameter function to model $\delta c^2/c^2$ , namely,
\begin{eqnarray}
\frac{\delta c^2}{c^2} = -\frac{A_{\delta}}{\Delta^2}\left(r-r_\rmd+\Delta\right)^2\left(\mathcal H\left[r-r_{\rmd}+\Delta\right]-\mathcal H\left[r-r_{\rmd}\right]\right),
\label{c2pert}
\end{eqnarray}
where $A_{\delta}$ and $\Delta$ are, respectively, the size of the jump and the width of the perturbation, $r_{\rm d}$ is the radius at which the jump in the sound speed takes place (i.e., the edge of the core) and $\mathcal H$ is the Heaviside Step Function \cite[e.g.][]{Abramowitz72}, with the argument shown in squared brackets. 

The right hand side (rhs) of eq.~(\ref{c2pert}) is shown in figure~\ref{c2}, by a continuous line. It is clear that this function provides an adequate fit to the perturbation derived from models. 


\subsection{Frequency perturbation\label{freqpert}}

When considering the perturbation to the oscillation frequencies resulting from the sharp variation of the sound speed at the edge of the core, we will restrict our analysis to linear, spherically symmetric, adiabatic oscillations. It is then possible to show that the perturbation to the oscillation frequency, $\delta\omega$, can be computed from the relative perturbation to the sound speed squared, through the relation (cf. eq~(B6) of Paper~I, expressed in terms of the independent variable $r$),
\begin{eqnarray}
2I_1\omega\delta\omega \approx -\int_{r_a}^{r_b}\frac{\rmd}{\rmd 
r}\left\{h\frac{\delta 
c^2}{c^2}\right\}\overline{\psi^2}\rmd r,
\label{dw1}
\end{eqnarray}
where $I_1$ is related to the mode inertia and $h$ is a function of $r$ that depends on the equilibrium properties of the star as well as on the frequency of the oscillations. These follow from the analysis presented in paper~I. Moreover, $r_{a}$ and $r_{b}$ define a bounded region where the perturbation to the sound speed is non-zero. Finally, $\overline{\psi^2} = \int\psi^2\rmd x$, where $\psi$ is related to the displacement eigenfunction and $x$ follows from a particular transformation of the independent variable $r$ (see paper~I for details).

Combining eqs~(\ref{c2pert}) and (\ref{dw1}) and making use of the properties of the Heaviside Step Function, we then find,
\begin{eqnarray}
2I_1\omega\delta\omega \approx  -A_\delta\left[h\overline{\psi^2}\right]_{r=r_\rmd}\hspace{-0.05cm}+\hspace{-0.05cm}\frac{A_{\delta}}{\Delta^2}\hspace{-0.05cm}\int_{r_{\rmd-\Delta}}^{r_\rmd}\hspace{-0.05cm}\frac{\rmd}{\rmd 
r}\left\{h\left(r-r_d+\Delta\right)^2\right\}\overline{\psi^2}\rmd r ,
\label{dw2}
\end{eqnarray}
where the squared brackets indicate that the function is to be taken at a particular radius, in the present case at $r=r_{\rm d}$. We note that the expression derived for the frequency perturbation in paper~I is equivalent to the expression that would be derived from eq.~(\ref{dw2}) if we were to keep only the first term on the rhs. However, in paper~I the amplitude $A_\delta$ was assumed to be proportional to the perturbation $\delta c^2 / c^2$ rather than equal to it. That arbitrariness no longer exists in the present analysis. As a result of the improvement in the modelling of  $\delta c^2 / c^2$, $A_\delta$ is, in the present case, exactly equal to the amplitude of the perturbation at the jump.

\begin{figure}
   \centering
   \includegraphics[width=8cm]{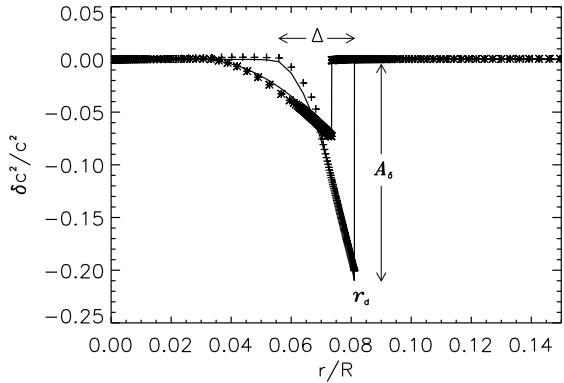}
      \caption{Relative sound speed squared perturbation for two models with $M=$1.3~M${_\odot}$ and ages of 2.25 Gyr (smaller amplitude) and 4.0 Gyr (larger amplitude). The symbols show the perturbation obtained by subtracting the squares of the sound speeds of the original and smooth models, and dividing by the the sound speed squared of the smooth model. The continuous lines show over-plots of the rhs of eq.~(\ref{c2pert}). Shown also are the quantities $A_\delta$, $\Delta$, and $r_{\rm d}$, that appear in eq.~(\ref{c2pert}), for the model of 4.0~Gyr of age.  }
         \label{c2}
   \end{figure}

\begin{figure}
   \centering
   \includegraphics[width=8cm]{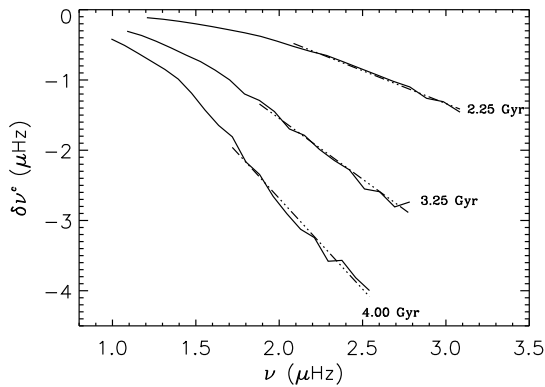}
      \caption{Perturbation to the cyclic oscillation frequencies $\delta\nu^c=(2\pi)^{-1}\delta\omega$,  induced by the sound speed perturbation at the edge of the core, as predicted by eq.~(\ref{dw2}). Shown are the results for the two models considered in figure~\ref{c2} and an additional model of the same mass with an intermediate age of 3.25 Gyr. The dotted-dashed lines show linear fits to the perturbations to the 10 modes of higher radial order.}
         \label{dw}
   \end{figure}

From eq.~(\ref{dw2}) we can see that the frequency perturbation is proportional to $A_\delta$, which implies that as the star evolves in the main sequence, and the amplitude of the sharp variation of the sound speed at the edge of the core increases, the amplitude of the frequency perturbation also increases.  This can be readily seen in figure~\ref{dw}, where the perturbation to the cyclic oscillation frequencies $\delta\nu^c=(2\pi)^{-1}\delta\omega$ computed from eq.~(\ref{dw2}) is shown for three models, including those considered in figure~\ref{c2}.  

Besides the overall increase in the perturbation, it is clear from figure~\ref{dw} that the derivative of the latter with respect to the oscillation frequency also increases with stellar age, being, in all cases, approximately constant at the higher end of the frequency range considered. The frequency ranges in which the perturbations have an approximately linear dependence on frequency are shown by the dotted-dashed lines in figure~\ref{dw}.

\begin{figure}
   \centering
   \includegraphics[width=8cm]{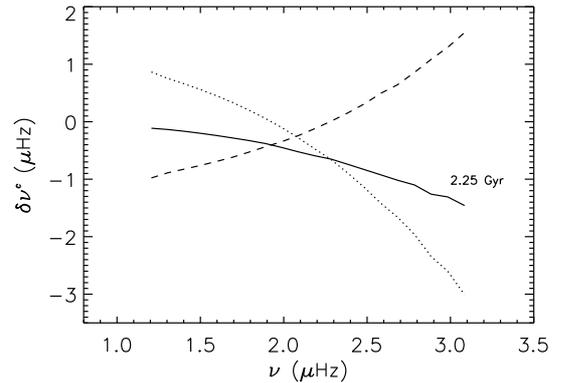}
   \includegraphics[width=8cm]{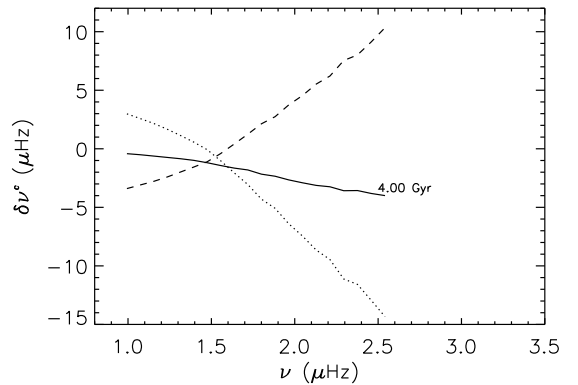}
      \caption{Same as in figure~\ref{dw} (continuous line) for the cases of models of 2.25 Gyr (upper panel) and 4.0 Gyr (lower panel). The dashed and dotted lines show, respectively, the contribution to the frequency perturbation of the first and second terms on the rhs of eq.~(\ref{dw2}).}
         \label{cont}
   \end{figure}

While the increase in the absolute value of the perturbation, as well as of its frequency derivative, with the age of the star is a prediction of the analysis, irrespectively of considering the first term only, or both terms on the rhs of eq.~(\ref{dw2}), we wish to note that the second term on the rhs is significantly smaller than the first term only when $\Delta$ is sufficiently large. Thus, generally speaking, both terms need to be taken into account in the computation of the frequency perturbation. This is illustrated in figure~\ref{cont}, for two of our models.






%

%

%

%

\section{Diagnostic tools\label{sec:quantities}}

\subsection{Small separations with modes of degree up to 3}

In order to constrain physical processes in the deepest layers of stars through asteroseismology it is necessary to explore the seismic data in a way such as to isolate particular signatures, like that derived in sec.~\ref{freqpert}. Since each oscillation frequency contains integrated information of the stellar interior, in this particular case that may imply combining different oscillation frequencies in a way such as to cancel all information they contain about the star, with the exception of the information about the sharp structural variation at the edge of the core. This, in turn, requires canceling out the information about the smooth component of the sound speed and the information about all sharp structural variations taking place in the stellar envelope. Such a diagnostic tool was proposed in Paper~I and we shall test it here, against the new functional form of the frequency perturbation, namely, that derived when including both terms on the rhs of eq.~(\ref{dw2}).

 The diagnostic tool proposed in Paper~I (hereafter named  $dr_{0213}$) combines modes of degree up to 3 according to the following expression,  
\begin{equation}
dr_{0213}\equiv\frac{D_{02}}{\Delta\nu_{n-1,1}}-\frac{D_{13}}{\Delta\nu_{n,0}}.
\label{tool}
\end{equation}
In the expression above, $D_{\ell,\ell+2}\equiv\left(\nu_{n,\ell}-\nu_{n-1,\ell+2}\right)/\left(4\ell+6\right)$ are the scaled small separations and 
$\Delta\nu_{n,\ell}\equiv\nu_{n+1,\ell}-\nu_{n,\ell}$ are the large separations, with $n$ and $l$ being, respectively, the mode radial order and angular degree. 
At high frequencies, this quantity has been shown to be related to the frequency perturbation by,
\begin{equation}
dr_{0213}\approx\frac{\delta\nu^c}{6\,{\Delta\nu_{n-1,1}}}.
\label{eq:signal}
\end{equation}

\begin{figure}
\includegraphics[width=8cm]{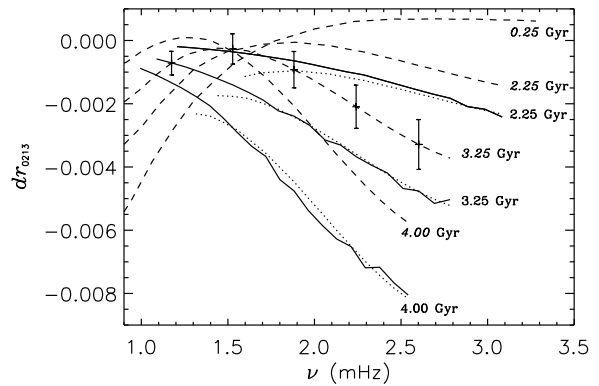}
\caption{Comparison between lhs (dashed lines) and rhs (full lines) of eq.~(\ref{eq:signal}) for the models shown in figure~\ref{dw2}. Error bars derived under the assumption of a relative error in the individual frequencies of $1\times 10^{-4}$ are shown for a selection of frequencies, for a model of 2.25 Gyr of age. The lhs of the same equation is shown also for a model of 0.25 Gyr, which has no convective core. The dotted lines show the lhs of eq.~(\ref{eq:signal}) for the three older models after subtracting a small constant. The value of the constant subtracted depends on the model age. In the present case these were $9.0\times 10^{-4}$, $1.5\times 10^{-3}$, and  $2.4\times 10^{-3}$ for, respectively, models of 2.25, 3.25, and 4.00 Gry of age.
 }
\label{signal}
\end{figure}

A comparison between the left and rhs of eq.~(\ref{eq:signal}) is shown in figure~\ref{signal} for the same three models of our evolution sequence considered before. We note that the rhs corresponds to the quantity shown in Fig~\ref{dw}, divided by six times the large separation.

If the diagnostic tool $dr_{0213}$ were capable of fully isolating the perturbation to the oscillation frequencies resulting from the sharp variation of the sound speed at the edge of the core, then the dashed and full lines for the three older models in figure~\ref{signal} would overlap. That is clearly not the case. To understand why, we show in the same figure the quantity $dr_{0213}$ for the youngest of our models, which does not have a convective core. This model without a convective core does not show a discontinuity in the sound speed in the innermost layers. Thus, one would expect $dr_{0213}$ to approach zero, at high frequencies. Instead, this frequency combination tends to a small constant. This is an indication that the information about the smooth component of the sound speed carried by the eigenfrequencies is not fully cancelled out when the latter are combined in the way to produce $dr_{0213}$. 

Despite the fact that  $dr_{0213}$ fails to fully isolate the signature of the sharp variation of the sound speed at the edge of the core, it is evident from figure~\ref{signal}, that at high frequencies the slopes of the two quantities are very similar. To verify this, we have subtracted small constants to the lhs of  eq.~(\ref{eq:signal}), and overplot the results in the same figure, using dotted lines. The dotted and full lines match well, confirming the similarity of the slopes. Since the frequency derivative of $\delta\nu^c$ is, at high frequencies, approximately constant (cf. figure~\ref{signal}) and proportional to $A_{\delta}$(cf. sec.~\ref{freqpert}), the diagnostic tool $dr_{0213}$ can still be used to extract information about the discontinuity of the sound speed at the edge of the core. In fact, that confirms and further explains the strong correlation found in Paper~I between the the slope of $dr_{0213}$ and $A_\delta$.

\subsection{Small separations with modes of degree up to 2}
It is well known that the oscillation frequencies of modes of degree $l=0$ or $l=1$ are the easiest to detect both in time-series photometry and spectroscopy. Modes of $l=2$ can still be detected if the data is sufficiently good. On the other hand, modes of $l=3$ are not generally detected in photometry and, while in principle they may be detected in ground-based spectroscopic data, very good data sets are needed to accomplish that.  
 
While the seismic diagnostic tool $dr_{0213}$ discussed above provides a way in which to measure the size of the discontinuity in the sound speed at the edge of the core, unfortunately one cannot expect to have the necessary data to apply it until precise and long spectroscopic time-series, such as those to be acquired by the network SONG \citep{grundahl09}, become available. Thus, in this section we turn our attention to diagnostic tools that may be applied to the data that are being acquired by space-based missions, which do not contain enough information about modes of $l=3$. 

Frequency combinations constructed with modes of degree up to 2 do not generally isolate the frequency perturbation originating from the edge of the core. However, if they involve modes of $l=0$, they must be affected by it. Thus, in what follows we will consider well known frequency combinations that involve only modes of degree smaller than 3 and investigate on how they are affected by the sharp structural variation at the edge of the core.

Let us, thus, consider, the quantity  \citep{gough83},
\begin{equation}
d_{02}=\nu_{n,\ell}-\nu_{n-1,\ell+2},
\label{d02}
\end{equation}
and the quantities \citep{roxburgh03},
\begin{eqnarray}
 d_{01}&=& 
\frac{1}{8}\left(\nu_{n-1,0} - 4\nu_{n-1,1}+6\nu_{n,0} - 4\nu_{n,1}+\nu_{n+1,0}\right) \nonumber\\
& & \\
d_{10} &=&
-\frac{1}{8}\left(\nu_{n-1,1} - 4\nu_{n,0}+6\nu_{n,1} - 4\nu_{n+1,0}+\nu_{n+1,1}\right).\nonumber
 \label{d010}
\end{eqnarray} 
We note that $d_{01}$ and $d_{10}$ are both 5-point combinations of modes of $l=0$ and 1, the first centered on modes of degree $l=0$ and the second centered on modes of $l=1$. In practice we will consider these two quantities together, denoting the result by $d_{010}$. For that, we compute the values of $d_{01}$ and $d_{10}$ for each $n$, and order them according to the central frequency, in a single set.

\begin{figure}
\includegraphics[width=8cm]{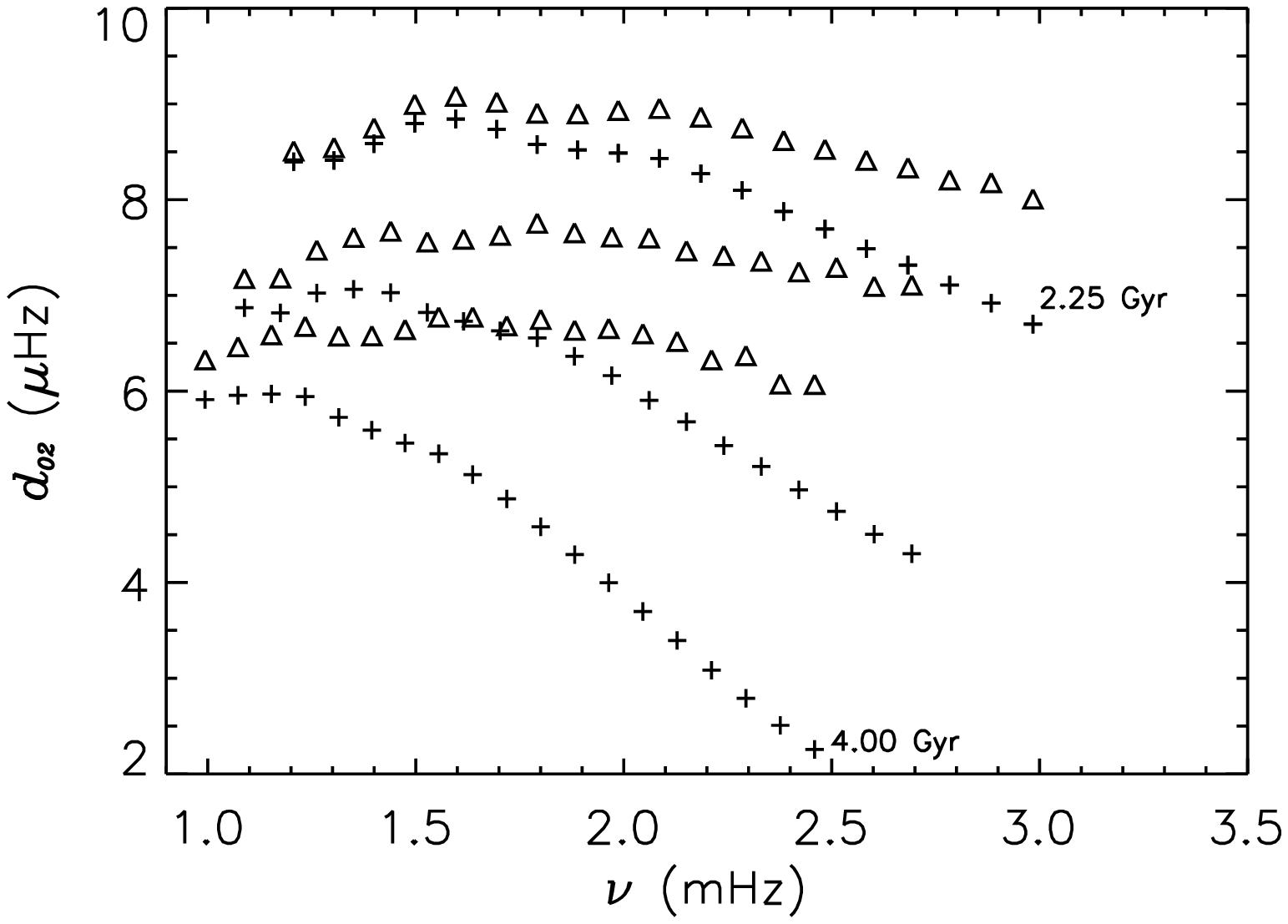}
\includegraphics[width=8cm]{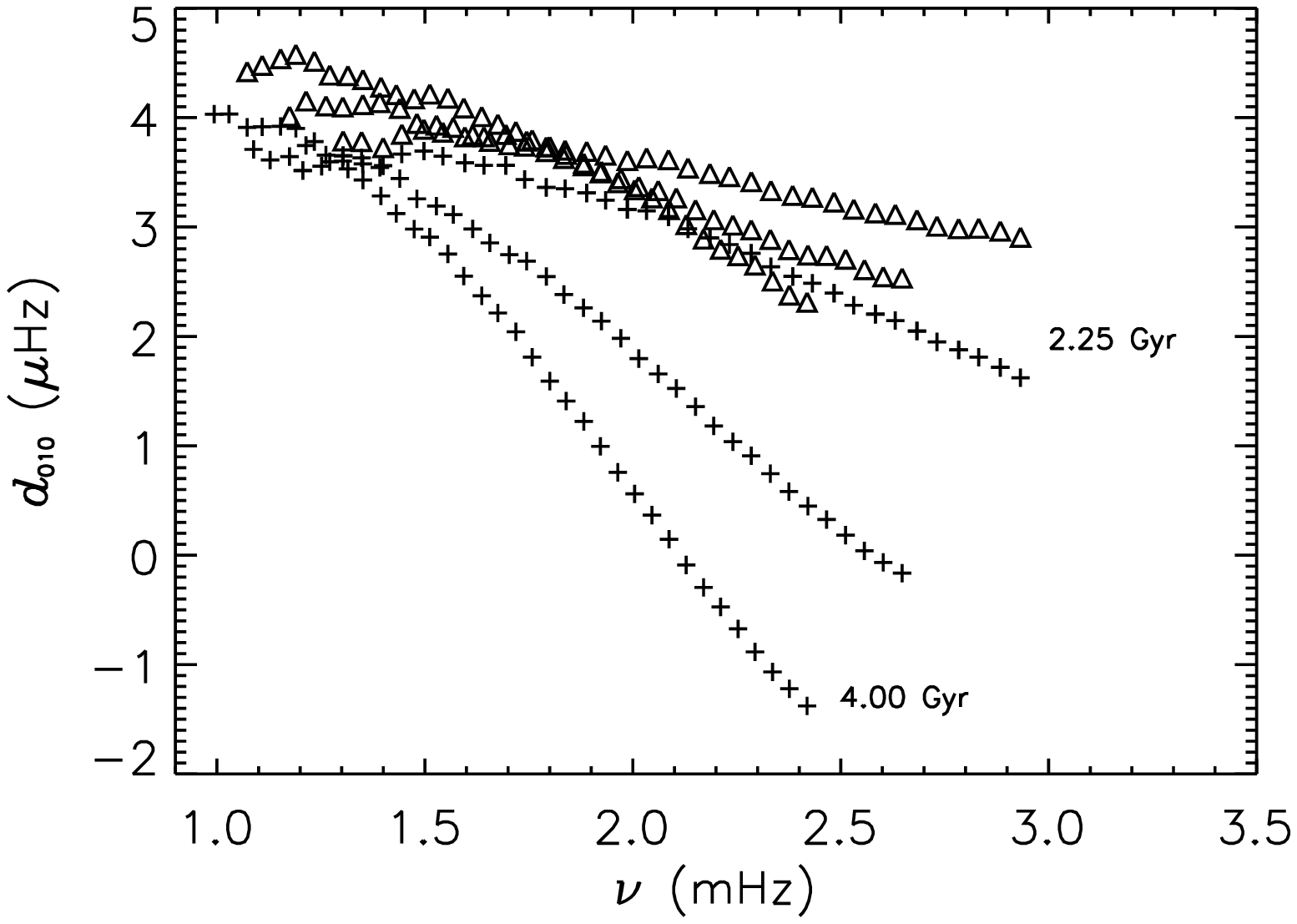}
\caption{Small separations, $d_{02}$ (upper panel) and $d_{010}$ (lower panel), shown as function of frequency, for the same three models as in Fig~\ref{dw}. Note that $d_{010}$ equals either $d_{01}$ or $d_{10}$, depending on whether the frequency against which it is plotted corresponds to a mode of $l=0$, or $l=1$, respectively. Crosses show these frequency combinations when derived from the eigenfrequencies of the original model, while open triangles show the same quantities when the perturbation to the frequencies of $l=0$ modes due to the sharp structural variation at the edge of the core is subtracted prior to the calculation of the frequency differences. 
 }
\label{d01d02}
\end{figure}

To investigate on the impact that the perturbation to the frequencies of $l=0$ modes has on the above mentioned frequency combinations, we have computed the latter for three models of our evolutionary sequence, in two different ways: 1) using the eigenfrequencies computed from the original equilibrium model; 2) using the same eigenfrequencies for modes of $l\ge 1$, but subtracting from the eigenfrequencies of $l=0$ modes the frequency perturbation derived from eq.~(\ref{dw2}).  Comparisons between the results obtained in these two cases are shown in figure~\ref{d01d02}. As expected, both frequency combinations are strongly affected by the sharp variation of sound speed at the edge of the core. In particular, it is clear from the figures that at high frequencies the frequency derivatives of these quantities increase significantly when the perturbation associated with the edge of the core is not subtracted. This is a direct consequence of the corresponding increase with frequency of $\delta\nu^c$.

\section{Discussion\label{sec:conclusions}}

In this work we have investigated on the effect that sharp structural variations taking place at the edge of convective cores of main-sequence solar-like pulsators have on the frequencies of radial oscillations, as well as on different small frequency separations constructed from the frequencies of low degree modes.

 Concerning the effect on the frequencies of radial oscillations, we found that the absolute value of the induced perturbation increases with frequency. Moreover, its frequency derivative, at high frequencies, is approximately constant, with an absolute value that increases with stellar age.  These results are in agreement with the findings of Paper~I and confirm that, if detected in the observations, these perturbations will provide information about the evolutionary status of the star.

Despite the qualitative agreement between our findings and those of Paper~I, we have shown that when the relative perturbation to the square of the sound speed at the edge of the core is modelled in line with the sound speed profiles seen in the stellar models considered, rather than by a simple step function, an additional, non-negligible term appears in the expression for the frequency perturbation. This results in the perturbation to the frequency of radial modes being negative at all frequencies, rather than changing sign at the frequency of the mode whose inner turning point coincides with the edge of the core, as had been found in Paper~I. Also, the frequency derivative of the perturbation is smaller when both terms are taken into account. 

In this work we have also considered the effect of the sharp structural variation at the edge of the core on three different small frequency separations.
With regards to the small separation proposed in paper~I, which involves modes of degree up to 3, we have shown that it is unable to fully isolate the frequency perturbation  $\delta\nu^c$, due to a small contamination originating from the smooth component of the sound speed. Nevertheless, we have confirmed that the frequency derivative of the same diagnostic tool can potentially be used to infer information about the amplitude of the ``discontinuity'' in the sound speed, hence, about stellar age.
Moreover, we have shown that the same structural variation induces a negative perturbation, that at high radial orders can amount to a few $\mu$Hz,  to the small frequency separations involving modes of $l=0,1$ and $l=0,2$, respectively. More importantly, the absolute values of the frequency derivatives of these small separations also increase significantly with the age of the star, as a consequence of the increase in the amplitude of the sharp variation in the sound speed at the edge of the core.

The dependence of  $\delta\nu^c$ on the oscillation frequency is a consequence of the frequency dependence of the location of the inner turning point of the modes. Moreover, the fact that its derivative is approximately constant at high frequencies is a consequence of the form of the pulsation eigenfunctions close to the same turning points.  If we were to consider modes of much larger radial orders (hence, even larger frequencies, for the same stellar model), we should eventually see the perturbation taking an oscillatory dependence on the oscillation frequency, as is seen when the sharp structural variation is placed well within the propagation cavity of the modes. We note, however, that modes of such high radial orders are not expected to be seen in these stars, due to them having frequencies above the acoustic cutoff frequency.  Therefore, for main-sequence solar-like pulsators, the main impact of the sharp structural variation at the edge of the core on both the oscillation frequencies of radial modes and the different small separations considered here is one of increasing the absolute value of their slopes when plotted against oscillation frequency, particularly at high radial orders. Most importantly, the increase in the slopes is highly dependent on stellar age.

Small separations constructed from modes of $l=0,1$ and, to a less extent, $l=0,2$ can in principle be determined from long photometric data-sets acquired from space. Thus, it is important to ask the question of whether the precision with which these quantities can be determined from the data is sufficient to detect the perturbations discussed in this work. Given that the perturbations are predicted to be of the order of  $\mu$Hz, we expect they might be seen in space-based data for some particular stars. However, there are a number of aspects that have to be taken into account when considering the question of detectability of this signature, such as stellar rotation, activity, and, of particular importance, mode lifetimes. An analysis of the detectability of the perturbation discussed in this work is undergoing, based on evolutionary sequences of stars of different masses and on realistic simulations of data.  That, will be presented in a separate paper.

\begin{acknowledgements}
 This work was supported by the project PTDC/CTE-AST/098754/2008 and the grant SFRH / BD / 41213 / 2007 funded by FCT/MCTES, Portugal. MC is supported by a Ci\^encia 2007 contract, funded by FCT/MCTES(Portugal) and POPH/FSE (EC).
\end{acknowledgements}

\bibliographystyle{aa}
\bibliography{solar-like} 

\begin{thebibliography}{22}
\expandafter\ifx\csname natexlab\endcsname\relax\def\natexlab#1{#1}\fi

\bibitem[{{Abramowitz} \& {Stegun}(1972)}]{Abramowitz72}
{Abramowitz}, M. \& {Stegun}, I.~A. 1972, {Handbook of Mathematical Functions},
  ed. M.~Abramowitz \& I.~A. Stegun

\bibitem[{{Arentoft} {et~al.}(2008){Arentoft}, {Kjeldsen}, {Bedding}, {Bazot},
  {Christensen-Dalsgaard}, {Dall}, {Karoff}, {Carrier}, {Eggenberger},
  {Sosnowska}, {Wittenmyer}, {Endl}, {Metcalfe}, {Hekker}, {Reffert}, {Butler},
  {Bruntt}, {Kiss}, {O'Toole}, {Kambe}, {Ando}, {Izumiura}, {Sato}, {Hartmann},
  {Hatzes}, {Bouchy}, {Mosser}, {Appourchaux}, {Barban}, {Berthomieu},
  {Garcia}, {Michel}, {Provost}, {Turck-Chi{\`e}ze}, {Marti{\'c}}, {Lebrun},
  {Schmitt}, {Bertaux}, {Bonanno}, {Benatti}, {Claudi}, {Cosentino}, {Leccia},
  {Frandsen}, {Brogaard}, {Glowienka}, {Grundahl}, \&
  {Stempels}}]{arentoftetal08}
{Arentoft}, T., {Kjeldsen}, H., {Bedding}, T.~R., {et~al.} 2008, \apj, 687,
  1180

\bibitem[{{Bedding} \& {Kjeldsen}(2007)}]{bedding07}
{Bedding}, T.~R. \& {Kjeldsen}, H. 2007, in American Institute of Physics
  Conference Series, Vol. 948, Unsolved Problems in Stellar Physics: A
  Conference in Honor of Douglas Gough, ed. {R.~J.~Stancliffe, G.~Houdek,
  R.~G.~Martin, \& C.~A.~Tout}, 117--124

\bibitem[{{Buzasi} {et~al.}(2000){Buzasi}, {Catanzarite}, {Laher}, {Conrow},
  {Shupe}, {Gautier}, {Kreidl}, \& {Everett}}]{buzasietal00}
{Buzasi}, D., {Catanzarite}, J., {Laher}, R., {et~al.} 2000, \apjl, 532, L133

\bibitem[{{Christensen-Dalsgaard}(2008)}]{cd08}
{Christensen-Dalsgaard}, J. 2008, \apss, 316, 13

\bibitem[{{Cunha} {et~al.}(2007){Cunha}, {Aerts}, {Christensen-Dalsgaard},
  {Baglin}, {Bigot}, {Brown}, {Catala}, {Creevey}, {de Souza}, {Eggenberger},
  {Garcia}, {Grundahl}, {Kervella}, {Kurtz}, {Mathias}, {Miglio}, {Monteiro},
  {Perrin}, {Pijpers}, {Pourbaix}, {Quirrenbach}, {Rousselet-Perraut},
  {Teixeira}, {Th{\'e}venin}, \& {Thompson}}]{cunhaetal07}
{Cunha}, M.~S., {Aerts}, C., {Christensen-Dalsgaard}, J., {et~al.} 2007,
  Astronomy and Astrophysics Review, 14, 217

\bibitem[{{Cunha} \& {Metcalfe}(2007)}]{cunha07}
{Cunha}, M.~S. \& {Metcalfe}, T.~S. 2007, \apj, 666, 413

\bibitem[{{Gilliland} {et~al.}(2010){Gilliland}, {Brown},
  {Christensen-Dalsgaard}, {Kjeldsen}, {Aerts}, {Appourchaux}, {Basu},
  {Bedding}, {Chaplin}, {Cunha}, {De Cat}, {De Ridder}, {Guzik}, {Handler},
  {Kawaler}, {Kiss}, {Kolenberg}, {Kurtz}, {Metcalfe}, {Monteiro}, {Szab{\'o}},
  {Arentoft}, {Balona}, {Debosscher}, {Elsworth}, {Quirion}, {Stello},
  {Su{\'a}rez}, {Borucki}, {Jenkins}, {Koch}, {Kondo}, {Latham}, {Rowe}, \&
  {Steffen}}]{gillilandetal10}
{Gilliland}, R.~L., {Brown}, T.~M., {Christensen-Dalsgaard}, J., {et~al.} 2010,
  \pasp, 122, 131

\bibitem[{{Gough}(1983)}]{gough83}
{Gough}, D.~O. 1983, Physics Bulletin, 34, 502

\bibitem[{{Grundahl} {et~al.}(2009){Grundahl}, {Christensen-Dalsgaard},
  {Arentoft}, {Frandsen}, {Kjeldsen}, {J{\o}rgensen}, \&
  {Kj{\ae}rgaard}}]{grundahl09}
{Grundahl}, F., {Christensen-Dalsgaard}, J., {Arentoft}, T., {et~al.} 2009,
  Communications in Asteroseismology, 158, 345

\bibitem[{{Houdek} \& {Gough}(2007)}]{houdek07}
{Houdek}, G. \& {Gough}, D.~O. 2007, \mnras, 375, 861

\bibitem[{{Michel} {et~al.}(2008){Michel}, {Baglin}, {Auvergne}, {Catala},
  {Samadi}, {Baudin}, {Appourchaux}, {Barban}, {Weiss}, {Berthomieu},
  {Boumier}, {Dupret}, {Garcia}, {Fridlund}, {Garrido}, {Goupil}, {Kjeldsen},
  {Lebreton}, {Mosser}, {Grotsch-Noels}, {Janot-Pacheco}, {Provost},
  {Roxburgh}, {Thoul}, {Toutain}, {Tiph{\`e}ne}, {Turck-Chieze}, {Vauclair},
  {Vauclair}, {Aerts}, {Alecian}, {Ballot}, {Charpinet}, {Hubert},
  {Ligni{\`e}res}, {Mathias}, {Monteiro}, {Neiner}, {Poretti}, {Renan de
  Medeiros}, {Ribas}, {Rieutord}, {Cort{\'e}s}, \& {Zwintz}}]{micheletal08}
{Michel}, E., {Baglin}, A., {Auvergne}, M., {et~al.} 2008, Science, 322, 558

\bibitem[{{Miglio} {et~al.}(2008){Miglio}, {Montalb{\'a}n}, {Noels}, \&
  {Eggenberger}}]{miglio08}
{Miglio}, A., {Montalb{\'a}n}, J., {Noels}, A., \& {Eggenberger}, P. 2008,
  \mnras, 386, 1487

\bibitem[{{Monteiro} {et~al.}(1994){Monteiro}, {Christensen-Dalsgaard}, \&
  {Thompson}}]{monteiroetal94}
{Monteiro}, M.~J.~P.~F.~G., {Christensen-Dalsgaard}, J., \& {Thompson}, M.~J.
  1994, A\&A, 283, 247

\bibitem[{{Monteiro} {et~al.}(2000){Monteiro}, {Christensen-Dalsgaard}, \&
  {Thompson}}]{monteiroetal00}
{Monteiro}, M.~J.~P.~F.~G., {Christensen-Dalsgaard}, J., \& {Thompson}, M.~J.
  2000, \mnras, 316, 165

\bibitem[{{Monteiro} \& {Thompson}(2005)}]{monteiro05}
{Monteiro}, M.~J.~P.~F.~G. \& {Thompson}, M.~J. 2005, \mnras, 361, 1187

\bibitem[{{Provost} {et~al.}(1993){Provost}, {Mosser}, \&
  {Berthomieu}}]{provost93}
{Provost}, J., {Mosser}, B., \& {Berthomieu}, G. 1993, \aap, 274, 595

\bibitem[{{Roxburgh} \& {Vorontsov}(1999)}]{roxburgh99}
{Roxburgh}, I.~W. \& {Vorontsov}, S.~V. 1999, in Astronomical Society of the
  Pacific Conference Series, Vol. 173, Stellar Structure: Theory and Test of
  Connective Energy Transport, ed. {A.~Gimenez, E.~F.~Guinan, \&
  B.~Montesinos}, 257--+

\bibitem[{{Roxburgh} \& {Vorontsov}(2001)}]{roxburgh01}
{Roxburgh}, I.~W. \& {Vorontsov}, S.~V. 2001, \mnras, 322, 85

\bibitem[{{Roxburgh} \& {Vorontsov}(2003)}]{roxburgh03}
{Roxburgh}, I.~W. \& {Vorontsov}, S.~V. 2003, \aap, 411, 215

\bibitem[{{Roxburgh} \& {Vorontsov}(2007)}]{roxburgh07}
{Roxburgh}, I.~W. \& {Vorontsov}, S.~V. 2007, \mnras, 379, 801

\bibitem[{{Walker} {et~al.}(2003){Walker}, {Matthews}, {Kuschnig}, {Johnson},
  {Rucinski}, {Pazder}, {Burley}, {Walker}, {Skaret}, {Zee}, {Grocott},
  {Carroll}, {Sinclair}, {Sturgeon}, \& {Harron}}]{walkeretal03}
{Walker}, G., {Matthews}, J., {Kuschnig}, R., {et~al.} 2003, \pasp, 115, 1023

\end{thebibliography}


\end{document}